\documentclass[prd, aps, a4paper, superscriptaddress, twocolumn,
nofootinbib]{revtex4}
\usepackage{graphicx}
\usepackage{color}
\usepackage{dcolumn}
\usepackage{bm}
\usepackage{slashed}
\usepackage{amsmath}
\usepackage{latexsym}
\usepackage{amssymb}
\usepackage{mathrsfs}
\usepackage{amsfonts}
\usepackage{url}
\usepackage[normalem]{ulem}
\usepackage{wasysym}
\usepackage[colorlinks]{hyperref}
\allowdisplaybreaks

\begin{document}
\title{Constraint on the fifth force through perihelion precession of
planets}
\author{Bing Sun}
\affiliation{Department of Astronomy, Beijing Normal University, Beijing
100875, China}
\author{Zhoujian Cao\footnote{corresponding author}} \email[Zhoujian Cao: ]{zjcao@amt.ac.cn}
\affiliation{Department of Astronomy, Beijing Normal University, Beijing
100875, China}
\author{Lijing Shao}
\affiliation{Kavli Institute for Astronomy and Astrophysics, Peking
University, Beijing 100871, China}

\begin{abstract}
The equivalence principle is important in fundamental physics. The fifth
force, as a describing formalism of the equivalence principle, may indicate
the property of an unknown theory. Dark
matter is one of the most mysterious objects in the current natural science.
It is interesting to constrain the fifth force of dark matter. We propose a
new method to use perihelion precession of planets to constrain the long-range fifth
force of dark matter. Due to the high accuracy of perihelion precession
observation, and the large difference of matter composition between the Sun
and planets, we get one of the strongest constraints on the fifth force of dark matter. In the near future, the
BepiColombo mission will be capable to improve the test by another factor of
ten.
\end{abstract}

\maketitle

\section{Introduction}

The equivalence principle (EP) is fundamental to both Newtonian theory and
Einsteinian theory \cite{liang00}. Consequently experimental examination of
the EP is very important to fundamental physics
\cite{Wagner_2012,Williams_2012,PhysRevLett.119.231101}. We have the
E\"otv\"os parameter to describe the violation of EP,
%--
\begin{align}
\eta^{(A,B)} \equiv \frac{2(a_A-a_B)}{a_A+a_B}, \label{equation1}
\end{align}
%--
where $a_i$ $(i=A,B)$ denotes the acceleration of two test particles $A$ and
$B$ relative to the central attractor. In general the attracting force arising from the
central attractor depends on the composition of two test particles and the
central object when EP is violated, so does the above E\"otv\"os parameter.
We need to distinguish between different E\"otv\"os parameters arising from
different objects. For example, (i) the MICROSCOPE satellite
\cite{PhysRevLett.119.231101} considered a Ti-Pt pair with respect to the
Earth, and obtained $\eta^{\rm (Ti,Pt)}_{\bigoplus}\lesssim10^{-14}$; (ii)
the E\"ot-Wash experiments \cite{Wagner_2012} considered Be-Al and Be-Ti
pairs with respect to the Sun, and obtained $\eta^{\rm
(Be,Al)}_{\bigodot},~\eta^{\rm (Be,Ti)}_{\bigodot}\lesssim10^{-13}$; and
(iii) Lunar laser ranging \cite{Williams_2012} considered the Earth-Moon pair
with respect to the Sun, and obtained
$\eta^{(\bigoplus,\leftmoon)}_{\bigodot}\lesssim10^{-13}$. Using the idea
proposed by \citet{PhysRevLett.70.119} and the above results
\cite{Wagner_2012,Williams_2012}, we have $\eta^{\rm (Be,Al)}_{\rm
DM},\eta^{\rm (Be,Ti)}_{\rm DM},\eta^{(\bigoplus,\leftmoon)}_{\rm
DM}\lesssim10^{-5}$ when considering the Galactic dark matter (DM) as the
attractor (see Refs.~\cite{adelberger2009torsion, Wagner_2012} for the framework of effective field theory).

Equivalently we can use the concept of fifth force to describe the violation
of EP~\cite{Burgess:1988qg}. Compared to the E\"otv\"os parameter, the concept of fifth force is more straightforward to be related to a fundamental theory.
Based on the fifth force, we can take the advantage of large difference in
the matter composition between two test bodies to better constrain the fifth
force of an object in question. This idea has been successfully applied
in Ref.~\cite{PhysRevLett.120.241104} where binary pulsar PSR J1713+0747
constrained a neutron star (NS)-white dwarf (WD) pair with respect to the DM,
$\eta^{\rm (NS,WD)}_{\rm DM}\lesssim0.004$ \cite{10.1093mnrassty2905}.

Dark matter is one of the most mysterious objects in current natural
science. It is interesting to study the fifth force behavior of dark matter
which will help people to advance knowledge about dark matter.
Due to the great efforts in searching for dark matter particles, we nowadays have stringent constraints on the interaction cross-section between ordinary matter and dark matter. Those studies mostly focus on the possible short-range interactions between dark matters and the nucleons. We here investigate another possibility with the {\it long-range} fifth-force formalism originally proposed by the authors of \cite{PhysRevLett.56.3}. Notice that the long-range fifth force we are studying is extremely weak from the point of view of particle physics. We will see that it is even weaker than the gravity interaction, thus it does not contradict any constraints from dark matter searches. This is a largely unexplored territory, thus it is interesting to see whether dark matter could have a sizeable long-range interaction with ordinary matter.
The strongest
constraint for the long-range fifth force of dark matter comes from the lunar laser
ranging (LLR) experiment. The NS-WD binary PSR
J1713+0747 also gives an interesting constraint using the large difference of
matter composition between NS and WD \cite{PhysRevLett.120.241104,10.1093mnrassty2905,Shao:2019vyk}. Here we propose a new method to use the
perihelion precession of planets to constrain the fifth force of dark matter.
Due to both the large difference of matter composition between the Sun and
the planets, and the high observational accuracy of perihelion precession, we
can get a very good constraint of the fifth force of the dark matter in the Galaxy.

The most well known system of celestial mechanics is the Sun-Mercury system.
The famous ``43 arcseconds'' problem hastened the birth of general
relativity. Later the observation about the Mercury perihelion precession
became more and more accurate \cite{PhysRevLett.120.191101}. The current
most accurate detection is done by the MESSENGER mission
\cite{genova2018solar}. In the near future much more accurate detection will
be achieved by the BepiColombo mission \cite{PhysRevLett.120.191101} which
was launched last year. These missions directly detect the relative distance
and velocity between Mercury and the Earth. Combining these data with
other related information, people can construct accurate ephemerides for
Mercury. Based on the accurate ephemerides it is straightforward to get the
perihelion advance of Mercury. Representative ephemerides for Mercury
include the JPL DE series \cite{folkner2014planetary}, the EPM series
\cite{Pitjeva2013} and the INPOP series \cite{verma2014use}. Take the EPM2004
ephemerides as an example; the estimated accuracy for the perihelion advance
of Mercury is about $10^{-3}$\,as/cy \cite{iorio2007first}. EPM2004 is
rather old; more recent ephemerides give much more accurate perihelion
advance of Mercury.

Although our knowledge is limited about the perihelion precession observation
for planets other than Mercury, we find that the Jupiter may result in a
better constraint than Mercury. This is because when one converts the
constraint of the extra perihelion precession to the E\"otv\"os parameter,
the planet orbit information plays an important role. We will calculate the
detailed conversion relation and explain such a dependence in the next
section. There an analysis of the E\"otv\"os parameter based on the
perihelion precession observation will also be presented. After that we
relate the E\"otv\"os parameter constraint to the fifth force in
Sec.~\ref{sec3}. At last, some related discussion and the summary are given
in Sec.~\ref{sec4}. To complement the main text, some necessary calculation
detail is included in the Appendix.

\section{Effects of the Fifth force on perihelion advance of the Mercury}

The fifth force results in a relative acceleration of the Mercury with
respect to the Sun~\cite{PhysRevLett.120.241104},
%--
\begin{align}
\vec{a}_{\eta_{\rm DM}} = \eta_{\rm DM} \vec{a}_{\rm DM}, \label{fifforce}
\end{align}
%--
where $\vec{a}_{\rm DM}$ is the gravitational acceleration acted on the
Mercury-Sun binary system by the DM in the Galaxy.
%--
\begin{figure}[htp]
\centering
\includegraphics[width = 0.5\textwidth]{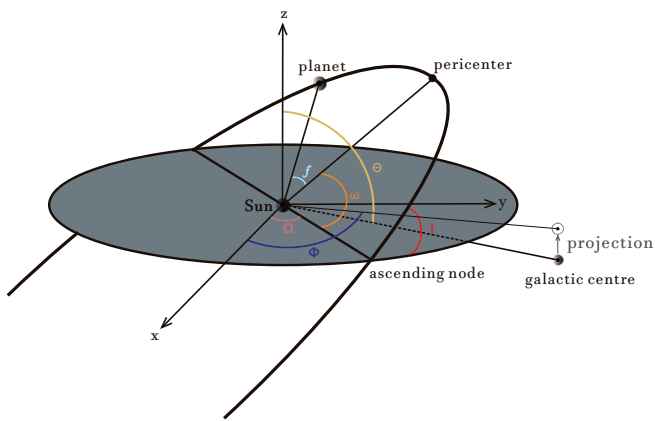}
\caption{The layout of the planet orbit with respect to the Galaxy center.}
\label{fig1}
\end{figure}
%--

The acceleration in Eq.~\eqref{fifforce} generates an additional perihelion
precession. This kind of physical picture has been investigated before by
other authors including \citet{PhysRevLett.66.2549} and \citet{Freire_2012}.
These authors were concerned about the orbital eccentricity variation. Differently
our concern is the precession of the perihelion. The variation of the longitude
of the perihelion can be expressed as (see the detailed calculation in
Appendix~\ref{app1}),
%--
\begin{align}
\dot{\varpi}_5 = &  - \frac{3\eta_{\text{DM}}a_{\text{DM}}\pi a^2}{GMP} \left(\frac{\sqrt{1-e^2}}{e}F_1+ \frac{e}{\sqrt{1-e^2}}F_2 \right), \label{varpi5}\\
F_1 = & \cos(\Phi-\Omega)\cos\omega\sin\Theta+\cos\Theta\sin\iota\sin\omega \nonumber \\
& + \sin\omega\cos\iota\sin\Theta\sin(\Phi-\Omega)\\
F_2 = & \tan\frac{\iota}{2} \sin\omega[\cos\iota \cos\Theta - \sin\iota \sin\Theta\sin(\Phi-\Omega)]
\end{align}
Here $G$ is the gravitational constant, $M$ is the total mass of the Sun and
the planet and $P$ is the orbital period. We have adopted traditional
notation above for the planet orbit, where $\Omega$ is the angle between the
$x$ direction and the ascending node, $\omega$ is the angle between the
perihelion and the ascending node of the ecliptic plane, $a$ is the
semi-major axis, $e$ is the orbit eccentricity, and $\iota$ is the
inclination of the orbit. In addition, $\Theta$ is the angle between the
Galactic center and the spin axis of the Sun, $\Phi$ is the angle between the
$x$ direction and the projected
direction of the Galactic center to the $x$-$y$ plane. The orbit layout is
illustrated in Fig.~\ref{fig1}. Specifically for the Galaxy center we have
$\Theta = 117.1^\circ$ and $\Phi = 192.9^\circ$.

The perihelion precession of planets is detectable. Until now the detected
perihelion precession of planets can be explained without the fifth force
contribution (\ref{varpi5}). So we can constrain $\eta_{\rm DM}$ through
Eq.~(\ref{varpi5}) based on the observational precision. Note that
Eq.~(\ref{varpi5}) can be viewed as,
\begin{align}
\dot{\varpi}_5 =\frac{\eta_{\rm DM}}{\Xi}
\end{align}
where the factor $\Xi$ is determined completely by the planet's orbit
information and the dark matter distribution. Based on the observational
precision $\epsilon_{\dot{\varpi}}$, we can roughly constrain the E\"otv\"os
parameter to,
\begin{align}
\eta_{\rm DM}<\Xi\cdot\epsilon_{\dot{\varpi}}.\label{eq3}
\end{align}

In order to determine the factor $\Xi$, we need to investigate the dark
matter distribution in our Galaxy. We
assume for the dark matter halo of our Galaxy,
\begin{align}
\rho &= \left\{
 \begin{array}{cc}
  \frac{\rho_{\rm sp}(r) \rho_{\rm in}(r) }{\rho_{\rm sp}(r) +\rho_{\rm in}(r) }\,, & r_0\leq r < R_{\rm sp}\\
   \rho_{\text{GNFW}}(r)\,, & r\geq R_{\rm sp}
   \end{array}\label{distr}
   \right.\\
\rho_{\text{GNFW}}(r) &= \frac{\rho_0}{(r/R_{s})^\gamma (1+r/R_s)^{3-\gamma}}. \label{GNFW}
\end{align}
The inner part of the halo is called the ``spike''. $\rho_{\rm sp} = \alpha
r^{-\gamma_{\rm sp}}$ is the distribution of the spike and $R_{\rm sp}$ is
its radius; $\gamma_{\rm sp} = \frac{9-2\gamma}{4-\gamma}$
\citep{ullio2001dark}. $\rho_{\rm in} = \beta r^{-\gamma_{\rm in}}$ has taken
into account the DM particles' annihilation cross section. $\gamma_{\rm in}$
depends on the annihilation mechanism: for $s$-wave annihilation $\gamma_{\rm
in} \simeq 0.5$, and for $p$-wave annihilation $\gamma_{\rm in} \simeq 0.34$
\cite{PhysRevD.93.123510}. The outer part is the generalized
Navarro-Frenk-White (GNFW) profile. Specifically for our galaxy $\gamma$
ranging from $1.0$ to $1.4$ \cite{PhysRevLett.120.241104} and $R_s = 2
0\,\text{kpc}$. Moreover, the parameter $\rho_0 = (2/7)^\gamma \times
3.2\times 10^{64} \, \text{GeV/kpc}^3 $ which is determined through the
condition that $\rho_{\text{GNFW}} (r\simeq 8\text{kpc}) = 1.2\times
10^{64} \, \text{GeV/kpc}^3$.

Integrating the density from $r_0=10^{-5} \, \text{kpc}$
\cite{PhysRevD.93.123510} to $R = 8 \, \text{kpc}$ where our Solar system
locates, we obtain the total DM mass,
\begin{align}
m_{\gamma} &= \int_{r_0}^R 4\pi r^2 \rho(r) {\rm d}r\\
&\in [8.4\times 10^{40} \, \text{kg},~1.0\times 10^{41} \, \text{kg}],
\end{align}
for $\gamma\in[1.0,1.4]$. We find that the spike part contributes little to
the total DM mass. So the parameter $R_{\rm sp}$ is negligible in the current
discussion \citep{PhysRevLett.120.241104}. Consequently, the gravitational
acceleration of DM can be estimated as,
\begin{align}
a_{\rm DM} = G\frac{m_{\gamma}}{R^2} \in [9.2 \times 10^{-11} \,\text{m/s}^2,~ 1.1\times 10^{-10} \,\text{m/s}^2]\,.\label{gac}
\end{align}
In the following analysis we take $a_{\rm DM}\approx10^{-10} \,\text{m/s}^2$.

Based on the above dark matter distribution information and planet orbit
information, we can determine the precession factor $\Xi$. Besides
Mercury we have also investigated several other planets, including Mars,
Jupiter and Saturn. We list these corresponding $\Xi$'s in
Table~\ref{tab1}.

\begin{table}[htbp]
\centering
\caption{Perihelion precession factor $\Xi$ of the Mercury, the Mars, the
Jupiter and the Saturn. Here the unit for $\Xi$ is (as/centrury)$^{-1}$.}
\label{tab1}
\begin{tabular}{p{1.8cm}<\centering |p{1.5cm}<\centering |p{1.5cm}<\centering |p{1.6cm}<\centering  |p{1.5cm}<\centering}
\hline
Planet & Mercury & Mars & Jupiter & Saturn\\
\hline
$\Xi$ & $0.271$ & $0.038$ & $0.0081$ & $0.031$\\
\hline
\end{tabular}
\end{table}

Currently the observational accuracy of perihelion precession of Mercury
is $10^{-3}$\,as/century \cite{iorio2007first,Park_2017,genova2018solar}. In
the near future, the European-Japanese BepiColombo mission will improve it
for Mercury to about $10^{-4}$\,as/century
\cite{BENKHOFF20102,PhysRevLett.120.191101}. Currently the observational
accuracy of perihelion precession of Saturn is $10^{-3}$\,as/century
\cite{Iorio_2009}. If more dedicated analysis is paid to construct an
ephemerides, a more accurate perihelion precession will be obtained
\cite{Pitjeva2005}. In the current paper we concern only the precession
directly deduced from experiments. Based on the current observational
accuracy, the E\"otv\"os parameter can be constrained to $\eta^{\rm
(Sun,Mercury)}_{\rm DM}<2.71\times10^{-4}$. In the near future the
BepiColombo mission will improve such a constraint to
$\eta^{\rm (Sun,Mercury)}_{\rm DM} \lesssim 3\times10^{-5}$. A comparable accuracy as
Mercury for Mars and Saturn will make the constraint one order of
magnitude better. For Jupiter it will make the constraint two orders of
magnitude better. But we are not sure about the observational accuracy of
perihelion precession for Mars, Jupiter and Saturn. In the
following analysis for the fifth force constraint we only discuss the result
from the Mercury observation.

\section{The Fifth Force of the Galactic DM}\label{sec3}
We assume that the fifth force between DM and ordinary matter can be
described by a Yukawa potential \cite{Wagner_2012,PhysRevLett.56.3},
\begin{align}
V(r) = \mp \frac{g^2}{4\pi} \frac{qq_{\rm DM}}{r}e^{-r/\xi}\label{eq2}
\end{align}
where $g$ is the coupling constant, $q_{\rm DM}$ and $q$ are the
dimensionless charges of DM and ordinary matter respectively, and $\xi$ is
the range of effective interaction which is related to the mass of the
intermediate particle through $\xi = \hbar/(mc)$. The $\mp$ sign corresponds
to scalar ($-$) and vector ($+$) interactions respectively. For an
electrically neutral body consisting of atoms, the charge $q$ is parametrized
as~\cite{Wagner_2012},
\begin{align}
q = Z\cos\psi + N\sin\psi \label{chartoratio}
\end{align}
where $Z$ is the proton number, $N$ is the neutron number, $\tan\psi \equiv
q_n/(q_p+q_e)$, and $q_p$, $q_n$, $q_e$ are the fifth force charge carried by
a proton, a neutron and an electron respectively. For a specific mixing, the value of $\psi$ can be derived; for example, in the $B-L$ scenario, $\psi = 90^\circ$.

The Yukawa potential (\ref{eq2}) gives rise to a relative acceleration of two
test bodies,
\begin{align}
\Delta a = \mp \frac{g^2}{4\pi}q_{\rm DM}\left( \frac{q_A}{m_A}-\frac{q_B}{m_B} \right)\left( \frac{1}{r^2}+\frac{1}{r\xi} \right)e^{-r/\xi},
\end{align}
where $q_{A,B}$ are the fifth force charges of these two bodies.
Notice that if we had replaced in the above equation the fifth-force charge $q_{\rm DM}$ with, say, $q_{\odot}$, the abnormal acceleration was tightly constrained by equivalence-principle experiments (e.g. by the E\"ot-Wash group and lunar laser ranging). Therefore, the fifth-force charges for ordinary matter are already constrained to be extremely small. In contrast, the fifth-force charges for dark matter are not so well bounded. In principle, dark matter can possess large fifth-force charges yet be un-noticed. This is exactly why it is interesting, and our study in this work is motivated.
Due to this relative acceleration, the EP will be violated with an E\"otv\"os parameter
\cite{Wagner_2012}
\begin{align}
	\eta_{\rm DM}^{(A,B)} &\approx\frac{\Delta a}{a_{\rm DM}} \nonumber \\
	&= \mp \frac{g^2}{4\pi G u^2} \left(\frac{q}{\mu} \right)_{\rm DM}\left[ \left(\frac{q}{\mu} \right)_A-\left(\frac{q}{\mu} \right)_B \right]\nonumber\\
&\times\left(1+\frac{r}{\xi} \right)e^{-r/\xi}, \label{DMeotvospar}
\end{align}
where $a_{\rm DM}$ is the acceleration resulted by the gravitational force of
the DM,
\begin{align}
a_{\rm DM} = \frac{Gm_{\rm DM}}{r^2},
\end{align}
and we have introduced atomic mass unit $u$ to express mass $m=u\mu$.
Based on the long-range interaction approximation to the fifth force
$\xi\rightarrow\infty$ \cite{PhysRevLett.120.241104}, the E\"otv\"os
parameter becomes,
\begin{align}
	\eta_{\rm DM}^{(A,B)} \simeq \mp \frac{g^2}{4\pi G u^2} \left(\frac{q}{\mu} \right)_{\rm DM}\left[ \left(\frac{q}{\mu} \right)_A-\left(\frac{q}{\mu} \right)_B \right] \,.
\end{align}
Using this expression, replacing the source of DM by some ordinary objects
such as the Earth, the Sun or various man-made objects, the EP violation test
can be formulated \cite{Wagner_2012}.

If we separate the acceleration of the ordinary material results from DM
into the gravitational part $a_{\rm DM}$ and the fifth force part
$a_{\eta_{\rm DM}}$, as $a_{\rm tol} = a_{\rm DM}+a_{\eta_{\rm DM}}$, the
E\"otv\"os parameter is related to the acceleration ratio through
\cite{PhysRevD.50.3614},
\begin{align}
	\frac{a_{\eta_{\rm DM}}}{a_{\rm tol}} =  \frac{\eta_{DM}^{(A,B)}\cos\psi}{\sin\psi\left[ \Delta(N/\mu)\right] + \cos\psi\left[\Delta(Z/\mu)\right]}\label{accratiodef}
\end{align}
where $\Delta(\cdot)$ means the difference between bodies A and B.
The acceleration ratio is determined up to an unknown $\psi$ once the
magnitude of $\eta_{\rm DM}$ and the compositions of matter in the experiment
are given.
%--
\begin{table}[htbp]
\centering
\caption{The ratios of proton number and neutron number to the
(dimensionless) mass for related objects in the current paper
\cite{adelberger2009torsion,PhysRevLett.120.241104}. Here the solid planets
mean planets like the Earth, Mercury, Mars and others.}
\begin{tabular}{p{2.7cm}<\centering p{2.7cm}<\centering p{2.7cm}<\centering}
\\
\hline
& Z/$\mu$ & N/$\mu$\\
\hline
NS & $0$ & $1.19$ \\
\hline
WD & $0.5$ & $0.5$  \\
\hline
solid planets & $0.49$ & $0.51$\\
\hline
Sun & $0.86$ & $0.14$\\
\hline
Moon & $0.502$ & $0.498$ \\
\hline
\end{tabular}
\label{ratiotable}
\end{table}

\begin{figure}[htp]
\centering
\includegraphics[width = 0.48\textwidth]{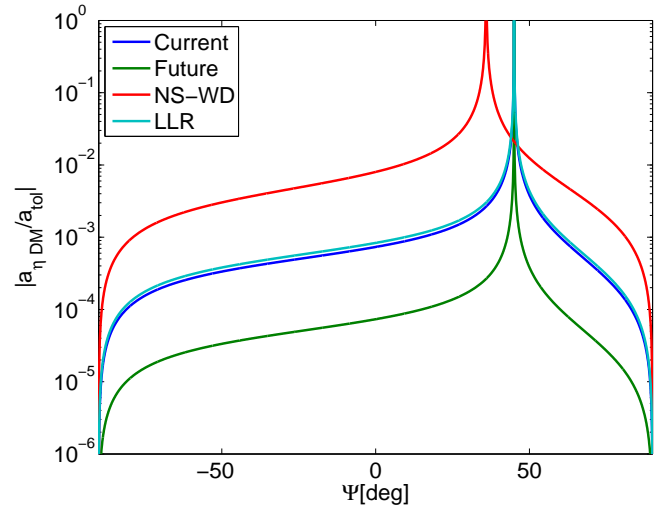}
\caption{The fifth force constraint  for neutral hydrogen from current perihelion precession measurement and
an expected near-future measurement based on the BepiColombo mission. For
comparison the constraints from the variation of orbital eccentricity of the
NS-WD binary PSR J1713+0747 \cite{PhysRevLett.120.241104} and that from the
LLR measurement \cite{Williams_2012,williams2009lunar} are also included.}
\label{fig2}
\end{figure}

We list the involved matter composition for kinds of objects in
Table~\ref{ratiotable}. Based on the current constraint $\eta^{\rm
(Sun,Mercury)}_{\rm DM}<2.71\times10^{-4}$ and a future expected constraint
$\eta^{\rm (Sun,Mercury)}_{\rm DM} \lesssim 3\times10^{-5}$ we can determine the
constraint of fifth force as shown in Fig.~\ref{fig2} for neutral hydrogen. In the figure, the regions above curves represent the excluded parameter space. For comparison we
reproduce the result for the NS-WD binary PSR J1713+0747 and that for LLR
\cite{PhysRevLett.120.241104}. Although the observational constraint of PSR
J1713+0747 on the E\"otv\"os parameter is not as tight $\eta^{\rm
(NS,WD)}_{\rm DM}\lesssim0.004$, the large difference of matter composition
makes the resulted fifth force constraint comparable to other experiments.
Regarding the LLR, although the observation accuracy makes the constraint to
the E\"otv\"os parameter very tight $\eta^{(\bigoplus,\leftmoon)}_{\rm
DM}\lesssim10^{-5}$, the similar matter composition for the Moon and the
Earth makes the constraint to the fifth force not quite outstanding among the
experiments. Interestingly our new perihelion precession method takes both
the advantage of high observation accuracy and a mediate matter composition
difference. As shown in Fig.~\ref{fig2}, the constraint resulted from current
observation is already similar to the result of LLR. In the near future one
more order of magnitude improvement will be achieved.
%--

\section{Discussion and Summary}\label{sec4}

The equivalence principle (EP) is important to gravitational theory and high
energy physics theory. Alternatively the equivalence principle can be
expressed as a fifth force. Since the equivalence principle, or the fifth
force, depends on the difference in the matter composition, it is useful to
investigate different kinds of matter. Due to the mysteries of dark matter and the possibility that dark matter might possess un-noticed large fifth-force charges,
it is quite interesting to investigate the equivalence principle related to
the dark matter from experiments and observations.

Regarding the dark matter in our Galaxy,
the most stringent constraint comes from the LLR detection. The authors of
Ref.~\cite{PhysRevLett.120.241104} took the advantage of the large matter
composition difference between a NS and a WD, and obtained a
compelling constraint.

In this paper we propose a new method to use the perihelion precession
observation of planets to constrain the fifth force acted by the dark matter
in our Galaxy. We interestingly find that
the E\"otv\"os parameter is proportional to an extra perihelion precession
rate. The proportional coefficient depends on, and only on, the gravitational
acceleration at the Solar system acted by dark matter and the orbit
information of the specific planet. Such dependence is shown in
Eq.~(\ref{varpi5}). Based on this relation, the E\"otv\"os parameter can be
constrained by the observational accuracy of the perihelion precession [see
Eq.~(\ref{eq3})]. For a given observation accuracy of the perihelion
precession, different planets result in different constraints on the
E\"otv\"os parameter. Due to the much smaller factor $\Xi$,
Jupiter gives two orders stronger constraint than Mercury if comparable accurate precession can be detected for Jupiter.

Thanks to the high observation accuracy of the perihelion precession, the current
constraint on the E\"otv\"os parameter has achieved $\eta^{\rm
(Sun,Mercury)}_{\rm DM}<2.71\times10^{-4}$. Thanks to the big difference of the matter composition between
planets and the Sun, the constraint on the E\"otv\"os parameter will result
in a good constraint on the fifth force. Besides our results, currently the
strongest constraint on the fifth force of the dark matter comes from the LLR. Current observation accuracy of the perihelion
precession for the Mercury results in a similar constraint. After the
BepiColombo mission, one more order of magnitude improvement in the
constraint is expected in the near future. It is worth mentioning that it is amazing to see that an extremely weak fifth force from the viewpoint of particle physics, even weaker than the gravity, can be constrained with celestial dynamics. This kind of study complements the searches for dark matter particles from, say, the Large Hadron Collider and underground laboratories.

\section*{Acknowledgments}
%|--------------------------------------------------------------------|
This work was supported by the NSFC (No.~11690023 and No.~11622546). Z. Cao was supported by ``the Fundamental Research Funds for the Central Universities.'' L.~Shao was supported by the Young Elite Scientists Sponsorship Program by the China Association for Science and Technology (2018QNRC001).
%|--------------------------------------------------------------------|

\appendix
\section{The effect of the Fifth force on the planet perihelion
precession}\label{app1}
We consider the fifth force as a perturbation to the binary motion. Then the
perturbative equations of the binary orbit elements are (see page 158 in
Ref.~\cite{poisson2014gravity})
\begin{align}
	\frac{da}{dt} = &2\sqrt{\frac{p^3}{\mu}}(1-e^2)^{-2}\left[-\sin f F_e+(e+\cos f)F_m\right] \label{at}\\
	\frac{d\iota}{dt} = & \sqrt{\frac{p}{\mu}}\frac{\cos(\omega+f)}{1+e\cos f}F_l \label{iotat}\\
	\sin\iota \frac{d\Omega}{dt} = & \sqrt{\frac{p}{\mu}}\frac{\sin(\omega+f)}{1+e\cos f}F_l\label{ascendt}\\
	\frac{de}{dt} = & \sqrt{\frac{p}{\mu}}\frac{1}{1+e\cos f} \bigg[  \left( 1+\cos^2f +2e\cos f \right) F_m\nonumber \\
	&  - \sin f\left( e+3\cos f +2\cos^2 f \right) F_e \bigg]  \label{et}\\	
	\frac{d\omega}{dt} = & \sqrt{\frac{p}{\mu}} \frac{1}{e(1+e\cos f)}\bigg[\frac{1}{2}(\cos 2f-2e\cos f-3)F_e \nonumber\\
	& + \sin f \cos f F_m -e \cot \iota \sin (\omega+f) F_l \bigg] \label{omegat}\\
	\frac{dM}{dt} = & n - \frac{1-e^2}{nae(1+e\cos f)}\bigg[ \big( 2 e \sin f+e \cos ^2f\nonumber \\
	 & -e \sin f \cos ^2f+2 \cos f-\sin f \cos f \big) F_e\nonumber \\
	 & +\big(-e \cos ^3f+2 e \cos f-e \sin f \cos f\nonumber \\
	 & -2 \sin f-\cos ^2f \big) F_m \bigg]\label{meant}\\
	\frac{df}{dt} = & \sqrt{\frac{\mu}{p^3}}(1+e\cos f)^2 \nonumber \\
	& + \sqrt{\frac{p}{\mu}} \frac{1}{e(1+e\cos f)}\bigg[ (2\sin f + \cos^2f \nonumber \\
	& +e\cos^3f+ e\sin f\cos f) F_e + (\cos f\sin f \nonumber \\
	& +e\cos^2f\sin f-2\cos f - e\cos^2f) F_m \bigg]\label{ft}
\end{align}
where $\mu = GM_{\rm tot}$, $p = a(1-e^2)$, $n=\sqrt{\mu/a^3}$, $F_e =
\vec{F}\cdot\hat{e}$, $F_m = \vec{F}\cdot\hat{m}$, $F_l =
\vec{F}\cdot\hat{l}$ with $M_{\rm tot}$ the total mass of the binary,
$\hat{e}$ the unit vector from the center of mass towards the perihelion,
$\hat{l}$ the unit vector pointing along the orbital angular momentum and
$\hat{m}=\hat{l}\times\hat{e}$.  $\vec{F}$ is the perturbed force.
$(a,~\iota,~\Omega,~e,~\omega,M,f)$ are the binary orbit elements,
respectively semimajor axis, inclination angle, ascending angle, periastron angle,
mean anomaly and true anomaly.

In order to investigate the secular change of the orbital elements, we
transform the derivatives with respect to $t$ to the ones with respect to the
true anomaly $f$,
\begin{align}
	\frac{da}{df} = & \frac{2p^3}{\mu}(1-e^2)^{-2}\nonumber\\
&\times\left[-\frac{\sin f}{(1+e\cos f)^2}F_e+ \frac{(e+\cos f)}{(1+e\cos f)^2}F_m\right] \label{af}\\
	\frac{d\iota}{df} = & \frac{p^2}{\mu} \frac{\cos(\omega+f)}{(1+e\cos f)^3}F_l \label{iotaf}\\
	\sin\iota \frac{d\Omega}{df} = & \frac{p^2}{\mu}\frac{\sin(\omega+f)}{(1+e\cos f)^3}F_l\label{ascendf}\\
	\frac{de}{df} = & \frac{p^2}{\mu} \frac{1}{(1+e\cos f)^3} \bigg[  \left( 1+\cos^2f +2e\cos f \right)F_m\nonumber \\
	&  - \sin f\left( e+3\cos f +2\cos^2 f \right)F_e \bigg]  \label{ef}\\
	\frac{d\omega}{df} = & \frac{p^2}{\mu} \frac{1}{e(1+e\cos f)^3}\bigg[ \frac{1}{2}(\cos 2f-2e\cos f-3)F_e \nonumber\\
	& + \sin f \cos fY -e \cot \iota \sin (\omega+f)F_l \bigg] \label{omegaf}\\
\frac{dM}{df} =& -\frac{p^2}{\mu}\frac{\sqrt{1-e^2}}{e(1+e\cos f)^3}\bigg[ \big( 2 e \sin f+e \cos ^2f\nonumber \\
	 & -e \sin f \cos ^2f+2 \cos f-\sin f \cos f \big)F_e\nonumber \\
	 & +\big(-e \cos ^3f+2 e \cos f-e \sin f \cos f\nonumber \\
	 & -2 \sin f-\cos ^2f \big)F_m \bigg].
\end{align}
For the Sun-planet binary systems with respect to the dark matter in the Galaxy, the fifth force can be
approximated as a constant force. Using this fact, and integrating the above equations with respect to
$f$ over the range $(0,2\pi)$, we obtain the change for a period of the binary
orbit. Combining this perturbation from the fifth force and the general
relativistic effect up to the first post-Newtonian order, the secular change
of the binary elements can be expressed as
\begin{align}
\left( \frac{da}{dt}\right)_{\text{sec}} = & 0\\
\left( \frac{d\iota}{dt}\right)_{\text{sec}} = & -\frac{3e}{2an\sqrt{1-e^2}}\cos\omega F_l  \\
\left( \frac{d\Omega}{dt}\right)_{\text{sec}} = & -\frac{3e}{2an\sqrt{1-e^2}}\frac{\sin\omega}{\sin\iota} F_l \\
\left( \frac{de}{dt}\right)_{\text{sec}} = & \frac{3}{2an}\sqrt{1-e^2}F_m \\
\left( \frac{d\omega}{dt}\right)_{\text{sec}}= & \frac{3\mu n}{c^2a(1-e^2)}-\frac{3}{2 an }\nonumber \\
&\times\left( \frac{\sqrt{1-e^2}}{e}F_e - \frac{e}{\sqrt{1-e^2}} \cot\iota\sin\omega F_l \right)\\
\left( \frac{dM}{dt}\right)_{\text{sec}} = & -\frac{n^3 a^2}{c^2e^2\sqrt{1-e^2}}\nonumber\\
&\times\left[ 5e^2+(6-7\eta)\left(1-\sqrt{1-e^2}\right) \right] \nonumber \\
& +\frac{1}{2na} \nonumber\\
&\times\bigg\{ \frac{2}{e^3}\left[ -1+3e^2+e^4+(1-e^2)^{5/2} \right]F_e \nonumber \\
&+(5-2e^2)F_m \bigg\} \\
\left( \frac{d\varpi}{dt}\right)_{\text{sec}}= & \frac{3\mu n}{c^2a(1-e^2)}-\frac{3}{2 an }\nonumber \\
&\times\left( \frac{\sqrt{1-e^2}}{e}F_e + \frac{e}{\sqrt{1-e^2}} \tan \frac{\iota}{2} \sin\omega F_l\right)\label{varpi5app}
\end{align}
where $\varpi = \omega+\Omega$. Equivalently we can use vector notation
$\vec{l} \equiv \sqrt{1-e^2} \hat{l}\,,~~\vec{e} \equiv e~\hat{e}$ to
denote the above relations
\begin{align}
&\left( d\vec{e}/dt\right)_{\text{sec}}= \frac{3}{2na}\vec{F}\times \vec{l} + \frac{3\mu n}{c^2a(1-e^2)}\vec{e}_z\times\vec{e},\\
&\left( d\vec{l}/dt \right)_{\text{sec}} = \frac{3}{2na}\vec{F}\times \vec{e}\,.
\end{align}
These two equations correspond to the second and the third expressions of
Eq.~(3) in \cite{PhysRevLett.66.2549}.

Converting to the variables involved in the main text, Eq.~\eqref{varpi5app}
reduces to Eq.~\eqref{varpi5}.

\bibliography{refs}

\end{document}